\documentclass[10pt,journal]{IEEEtran}

\usepackage{amsthm, amssymb}

\def\log{\textrm{log}}

\newtheoremstyle{slplain}
  {3pt}
  {3pt}
  {\slshape}
  {}
  {\bfseries}
  {.}%
  { }
  {}

\theoremstyle{slplain}

\usepackage{cite}
\usepackage{amsfonts}
\usepackage{multicol,multienum}
\usepackage{subfigure}
\usepackage{multirow}

\ifCLASSINFOpdf
  \usepackage[pdftex]{graphicx}
  \graphicspath{{../pdf/}{../jpeg/}}
  \DeclareGraphicsExtensions{.pdf,.jpeg,.png}
\else
  \usepackage{float}
  \usepackage[dvips]{graphicx}
  \graphicspath{{../eps/}}
  \DeclareGraphicsExtensions{.eps}
\fi

\usepackage[cmex10]{amsmath}
\usepackage{algorithm}
\usepackage{algorithmic}
\usepackage{array}
\usepackage{url}
\usepackage{setspace}

\begin{document}

\title{Dynamic Spectrum Refarming\\ with Overlay for Legacy Devices}

\author{
Xingqin Lin and Harish Viswanathan
\thanks{Xingqin Lin is with the Department of Electical and Computer Engineering at The University of Texas at Austin, Austin, TX. Harish Viswanathan is with Bell Labs, Alcatel-Lucent, Murray Hill, NJ. This work was performed when Xingqin was a summer intern at Bell Labs. Email: xlin@utexas.edu, harish.viswanathan@alcatel-lucent.com. Date revised: \today.}
}

\maketitle

\begin{abstract}
The explosive growth in data traffic is resulting in a spectrum crunch forcing many wireless network operators to look towards refarming their 2G spectrum and deploy more spectrally efficient Long Term Evolution (LTE) technology. However, mobile network operators face a challenge when it comes to spectrum refarming because 2G technologies such as Global System for Mobile (GSM) is still widely used for low bandwidth machine-to-machine (M2M) devices. M2M devices typically have long life cycles, e.g. smart meters, and  it is expensive to migrate these devices to newer technology since a truck roll will typically be required to the site where a device is deployed. Furthermore, with cost of 2G modules several times less than that of LTE, even newly deployed M2M devices tend to adopt 2G technology. Nevertheless, operators are keen to either force their 2G M2M customers to migrate so that they can refarm the spectrum or set aside a portion of the 2G spectrum for continuing operating 2G and only refarm the rest for LTE. In this paper we propose a novel solution to provide GSM connectivity within an LTE carrier through an efficient overlay by reserving a few physical resource blocks for GSM. With this approach, operators can refarm their 2G spectrum to LTE efficiently while still providing some GSM connectivity to their low data rate M2M customers. Furthermore, spectrum can be dynamically shared between LTE and GSM. An approach similar to that proposed in this paper can also be applied for other narrow band technology overlays over LTE.
\end{abstract}

\IEEEpeerreviewmaketitle


\section{Introduction}


The exponentially growing demand for wireless traffic is straining cellular networks.  According to Cisco's report \cite{cisco2011cisco}, in 2011 global mobile data traffic doubled for the fourth year in a row and was eight times the traffic of the entire global Internet in 2000. Cisco further expects that global mobile data traffic will increase 18-fold between 2011 and 2016 \cite{cisco2011cisco}. Obviously, mobile network operators are facing intense pressure to make their networks keep pace with this exponential wireless traffic growth.

The explosive data growth is resulting in a spectrum crunch and thus many operators are looking to refarm their 2G spectrum to deploy more spectrally efficient Long Term Evolution (LTE) technology. LTE is the key standard for wireless communications providing high speed data services in next-generation cellular networks \cite{website:3gppLTE, Astely2009LTE, Ghosh2010Fundamentals}. Not surprisingly, evolution to LTE is critical for mobile network operators to deliver the high speed mobile broadband services that their customers demand. However, operators face a challenge when it comes to spectrum refarming because of the numerous legacy machine-to-machine (M2M) devices \cite{Vodafone2010M2M, Ericsson2011M2M}.  M2M devices typically have long life cycles, e.g. smart meters, and it is expensive to migrate these devices to LTE since a truck roll is typically required to the site where a device is deployed.

Meanwhile, GSM is - and will remain attractive - for M2M. GSM is particularly suitable for M2M because most M2M devices require low data rate communication that is adequately met by GSM at module prices that are less than a fifth of that of third and fourth generation modules. Furthermore, because GSM is available worldwide \cite{mehrotra1997gsm} a single technology device based on GSM can be used for applications deployed internationally. In fact, two thirds of all new cellular M2M modules shipped in 2011 are estimated to be GSM/GPRS by ABI Research \cite{GSMModules-ABI}.

The above discussions imply that mobile network operators may have to keep providing GSM service for legacy devices, especially M2M, although they would like to refarm their GSM spectrum for LTE. In this paper we propose a novel  dynamic spectrum refarming (DSR) approach  for the co-existence of GSM and LTE. DSR allows complete refarming of the GSM spectrum for LTE. But some LTE physical resource blocks (PRB) will be reserved for GSM transmission, i.e., LTE Evolved Node B (eNodeB) will not schedule those reserved PRBs for any User Equipment (UE) and accordingly suppress the reference signals.  With this approach, operators can migrate their GSM spectrum to LTE while still providing GSM connectivity to their low data rate M2M customers. More importantly, spectrum can be dynamically shared between LTE and GSM simply through adjusting the number of reserved PRBs. This approach is advantageous compared to static partitioning of the legacy spectrum into a portion for GSM and a portion for LTE because of more efficient use of spectrum. In addition, the basic idea of reserving LTE PRBs can be applied to other scenarios as well; for example, similar idea is used in \cite{lin2013dynamic} for deploying LTE small cells sharing the same spectrum as existing GSM networks.

Despite the potential of the proposed GSM overlay on LTE for legacy devices, there are several technical issues to be addressed. First and foremost, will LTE UEs with no modifications continue to operate as usual? This is not \textit{a priori} clear since reserving PRBs for GSM modifies the channel structure of LTE. Secondly, will GSM mobiles operate properly? This is also not \textit{a priori} clear since GSM PRBs are adjacent to LTE PRBs  and the LTE interference leakage may cause problems. If so, what are the corresponding solutions? Last but not least, is the impact on the LTE spectral efficiency because of the GSM overlay significant? What are the methods to recover the loss of LTE capacity because of adjacent channel interference from GSM?  We address these critical issues in this paper.

The rest of this paper is organized as follows. Section \ref{sec:basic} describes the basic idea and the advantages of DSR by a specific example. We then discuss how to reserve PRBs for GSM and mitigate the inter-technology interference between LTE and GSM in Section \ref{sec:overcome}. Several LTE enhancements to recover the capacity loss  are presented  in Section \ref{sec:lte}. Section \ref{sec:sim} presents some simulation results, and is followed by the discussion on uplink design in Section \ref{sec:uplink} and the conclusions in Section \ref{sec:conclusion}.

\section{Basic Idea and Benefits of DSR}
\label{sec:basic}

\begin{figure}
 \centering
 \includegraphics[width=8cm]{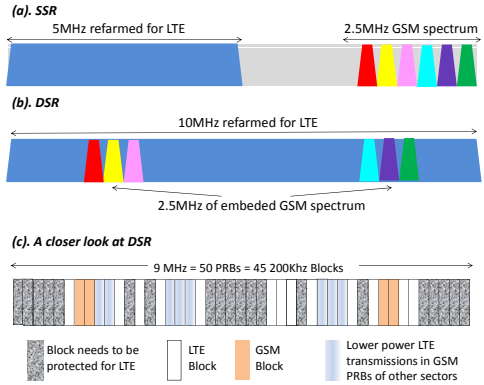}
 \caption{Static spectrum refarming versus dynamic spectrum refarming. Note that the bottom subfigrue plots 45 200KHz blocks rather than 50 180KHz LTE PRBs. As GSM center frequency must be an integer multiple of 200KHz \cite{ETSIgsm}, GSM channel edges may not be always aligned with LTE PRB edges.}
 \label{fig:staticVSdynamic}
 \end{figure}

In this section, we use a specific example to illustrate the basic idea and benefits of DSR. Suppose that a mobile network operator has 10MHz spectrum in the GSM band and that the capacity requirement of GSM has diminished so that 2.5MHz is sufficient, as shown in Fig.\ref{fig:staticVSdynamic} (a) and (b).

\subsection{DSR: A high level description}

The basic idea of DSR is to exploit the flexibility of OFDM used in LTE to embed GSM transmissions within a portion of LTE transmission. In particular, with DSR an LTE eNodeB reserves certain PRBs within the 10MHz LTE carrier for GSM and does not transmit any LTE signal on those subcarriers.\footnote{Note that, with 10MHz LTE transmission bandwidth, the actual occupied bandwidth is 9MHz while the guard band consumes 1MHz bandwidth.} Instead, GSM signals are transmitted from the same base station on the reserved PRBs.

Obviously, PRBs that are reserved for GSM need to be carefully picked so that the critical LTE PRBs used for synchronization, control signaling and other signaling such as hybrid automatic repeat request (HARQ) feedback are not allowed for use by GSM. With this approach, LTE UEs are not significantly impacted by GSM transmissions. The adjacent channel leakage ratio of GSM is such that the SINR of neighboring LTE PRBs will be limited to at most $11$dB because of GSM signal interference assuming that GSM power spectral density (PSD) is $13.56$dB higher\footnote{GSM PSD is assumed to be $13.56$dB higher than LTE PSD as typically $20$W transmit power is used per $200$KHz GSM channel while $40$W transmit power per $9$MHz LTE channel.} than that of LTE. (The $11$dB cap on LTE SINR can be calculated using the GSM power leakage profile specified in Table \ref{tab:gprs} and plotted  in Fig. \ref{fig:gprs}.) However, this does not significantly impact overall LTE spectral efficiency because LTE UEs close to cell edge already have SINR limited to less than $1$dB because of out-of-cell interference. To mitigate interference from LTE PRBs to GSM, transmit power on the PRBs close to GSM PRBs can be reduced. Finally, GSM requires frequency reuse. Spectral efficiency can be improved by allowing low power LTE transmissions to LTE UEs close to the base station on the GSM PRBs of the neighboring cells/sectors. With this fractional reuse approach between LTE and GSM, the amount of spectrum needed to support GSM can be minimized. With the above techniques an overlay can be supported efficiently.

The above high level description on DSR will be explored more in later sections. Here we show an example allocation of GSM spectrum with LTE in Fig.\ref{fig:staticVSdynamic}(c). Each block in the figure represents 200KHz, the width of a GSM carrier. A total of 12 200KHz blocks or about 2.4MHz of spectrum can be assigned as default GSM spectrum. This represents about 25\% of the LTE spectrum. Each sector can have a GSM Broadcast Control Channel (BCCH) carrier and a traffic carrier. A frequency reuse of 3/9 can be supported for BCCH and 1/3 for traffic channel. As an alternative to frequency hopping, antenna selection diversity can be employed where the signal is transmitted on different antennas in alternate slots to provide diversity transmission. The picture also shows  that some of the GSM blocks meant for use in other cells can be used in this cell for LTE with low power so that they do not cause interference to other cells. The grey blocks cannot be assigned to GSM since they need to be protected for LTE control signaling. Detailed descriptions can be found in the next section. The spectrum sharing can be dynamic in the sense that when there is no GSM traffic in a given sector, the GSM blocks can be used for LTE transmission.

Before ending this subsection, it should be noted that the number of PRBs that are reserved for GSM should be appropriately determined: there would be inefficiency or loss if either excess or too few PRBs are reserved. In DSR, the number of reserved PRBs is based on relatively long term statistics such as hours. In contrast, the durations of GSM sessions are at the time scale of seconds or at most minutes. Thus, on a short time scale the reserved PRBs may be under utilized or over loaded. Considering that M2M traffic such as meter reading application is typically delay tolerant, M2M traffic on GSM can be time-shifted to avoid over-loading. In this way, the reserved PRBs utilization can be made efficiently even though the number of reserved PRBs are determined based on long term average GSM traffic trends.

\subsection{DSR versus SSR}

We compare our proposed DSR to static spectrum refarming (SSR) to highlight the key advantages of DSR. SSR completely separates GSM and LTE within the band, i.e., part of the GSM spectrum is kept for legacy devices while the remaining GSM spectrum is refarmed for LTE. As a specific example, consider the scenario shown in Fig.\ref{fig:staticVSdynamic}. Because LTE transmission bandwidth can only be 1.4, 3, 5, 10, 14, or 20 MHz wide, the operator will be restricted to refarming 5 MHz of the spectrum for LTE and the remaining 2.5MHz  will be under utilized. Even if LTE can simultaneously support a 5MHz channel and a 1.4MHz channel by e.g. carrier aggregation (see, e.g., \cite{Iamura2010CA, Shen2012over, lin2012modeling}), the remaining 1.1MHz spectrum is under utilized. In contrast, DSR allows the deployment of 10MHz LTE channel with 2.5MHz of GSM embedded. Thus, compared to SSR, DSR can minimize the wastage of spectrum.

The second advantage of DSR is its flexible spectrum reuse capability. In SSR, the LTE bandwidth is fixed, e.g., 5MHz in Fig.\ref{fig:staticVSdynamic}. The system cannot reassign bandwidth between GSM and LTE with changing traffic demand. In contrast, with DSR LTE in a given sector can utilize the GSM carriers of the neighboring sector with low transmit power. Low power LTE transmissions targeting good geometry users can provide significant spectral efficiency. Since transmissions are of low power,  GSM transmissions in the neighboring cell are (nearly) not affected. This idea can be viewed as a form of fractional frequency reuse (FFR) (see, e.g.,\cite{Boudreau2009Interference, Stolyar2008Self, Ali2009ffr}) for inter-technology inter-cell interference coordination (ICIC). Besides, if immediate transmission is not strictly required for M2M such as for meter reading application, M2M traffic on GSM can be scheduled to avoid the busy hour and DSR allows more spectrum to be used for LTE to accommodate the needs.\footnote{Wireless traffic is imbalanced over both time and space. During the daytime, the demand is high in working areas and is low in the residential area. The converse is true during the night.}

To sum up, DSR provides GSM connectivity within an LTE carrier through an efficient, dynamic overlay by reserving a few PRBs for GSM.

\section{Overcoming System Design Issues}
\label{sec:overcome}

\subsection{How to Reserve PRBs}

DSR solution reserves certain PRBs within the LTE transmission bandwidth for GSM signals and does not transmit any LTE signal on those subcarriers. This reservation scheme changes the channel structure of LTE. To ensure normal operation of LTE, we need to carefully select the positions of those reserved PRBs out of the LTE channel. We next take the $10$MHz LTE channel as an example and discuss how the system should reserve PRBs to avoid/minimize the potential impact of LTE puncturing on the broadcast and synchronization channels, reference symbols, and other control signaling channels including Physical Control Format Indicator Channel (PCFICH), Physical Downlink Control Channel (PDCCH) and Physical Hybrid-ARQ Indicator Channel (PHICH).

\subsubsection{Avoid broadcast and synchronization channels}
\begin{figure}
 \centering
 \includegraphics[width=8cm]{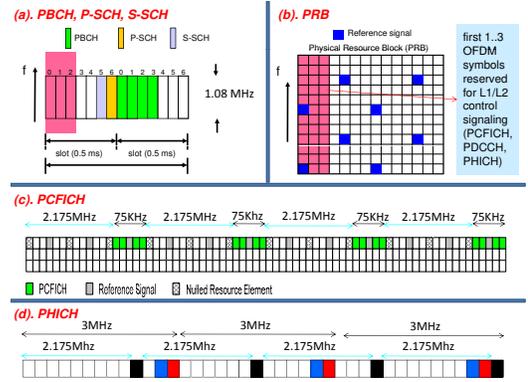}
 \caption{LTE downlink channel structure}
 \label{fig:resource}
 \end{figure}

LTE broadcast and synchronization channels are located within the central $1.08$MHz of the $10$MHz LTE channel, as shown in Fig.\ref{fig:resource}(a). So we can avoid reserving the PRBs used by the LTE broadcast and synchronization channels.

 \subsubsection{No impact on reference signals}
 The impact of puncturing on reference symbols is not an issue because the reserved PRBs will not be scheduled for any user and thus the corresponding reference signals are suppressed.

 \subsubsection{Impact on PCFICH and PHICH}
 The PCFICH occupies four 75KHz chunks within the $10$MHz LTE channel, as shown in Fig.\ref{fig:resource}(c). As for PHICH, multiple PHICHs are mapped to the same set of resource elements and these PHICHs constitute a PHICH group. The number of PHICH groups can be configured by higher layers to some extent. However, the positions of PCFICH and PHICH are not fixed and vary with the physical cell ID \cite{3gppMobility}. They would spread out and occupy all the PRBs  if all the cell IDs were used and we would not be able to reserve PRBs without affecting them. To overcome this issue, we can only use a small subset of the cell IDs. Then PCFICH and PHICH will only occupy certain part of the LTE transmission bandwidth, which can be avoided by the reserved PRBs.

  \subsubsection{Impact on PDCCH}
 The real challenge comes from the PDCCH which occupies all the PRBs. This implies that a certain number of resource element groups carrying the PDCCH have to be wiped out with LTE puncturing. Fortunately, LTE multiplexes and interleaves several PDCCHs within one LTE subframe. This helps to spread the impact of puncturing over all the PDCCHs and each PDCCH only needs to tolerate a certain level of the errors. Moreover, LTE allows the system to increase the aggregation level of the control channel elements, which can make PDCCHs more robust against errors.  Note that given limited resource of the control channel elements, increasing its aggregation level reduces the number of PDCCHs that can be simultaneously used.

\begin{figure}
\centering
\includegraphics[width=8cm]{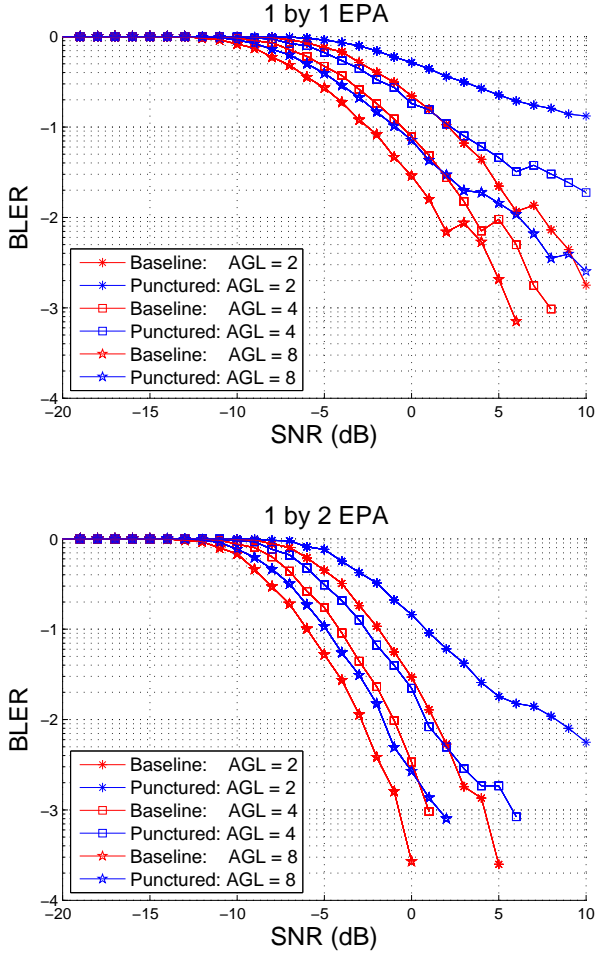}
\caption{Impact of puncturing on PDCCH: 10MHz LTE channel with a 2.4MHz punctured portion. LTE EPA multipath fading channel model is used. AGL denotes the aggregation level of the control channel elements.}
\label{fig:pdcch}
\end{figure}

We further perform link level simulation to study the impact of puncturing on PDCCH. The result for LTE Extended Pedestrian A (EPA) model is shown in Fig. \ref{fig:pdcch}. As expected, the PDCCH block error rate (BLER) increases due to the puncturing. However, the loss in BLER can be recovered by increasing one aggregation level of the control channel elements. Besides, $1 \times 2$ system has better BLER performance than $1 \times 1$ system, which is natural since the former provides higher diversity order. Note that in Fig. \ref{fig:pdcch} we assume that GSM PSD is   as high as LTE PSD in the link simulation but DSR is transparent to LTE receivers (i.e., LTE receivers do \textit{not} know certain portions of the channel have been punctured for GSM overlay). We notice that PDCCH under DSR has unacceptably high BLER if  GSM PSD is set to be $13.56$dB higher than LTE PSD and LTE receiver includes all PDCCH symbols in the decoder. In constrast, if LTE receivers take into account certain portions of the channel have been punctured for GSM overlay when decoding  PDCCH, then PDCCH under DSR has better BLER performance than the performance of its counterpart presented in Fig. \ref{fig:pdcch}. We do not present the specific numerical results here due to space constraints.

The awareness of GSM signals at LTE UEs can be achieved by explicitly signaling to the LTE UEs. The information about spectrum puncturing can be conveyed together with other system information in the Physical Broadcast Channel (PBCH), which is not affected by the proposed puncturing. Explicit signaling may be avoided at the cost of additional complexity in the receiver design of LTE UEs. For example, the LTE receivers can detect the presence of GSM through energy detection. This is possible because those symbols corresponding to the GSM PRBs will be received with much higher power relative to the other LTE symbols.


\subsubsection{Impact on channel estimation quality} With the awareness of the presence of GSM overlay (required to achieve acceptable PDCCH performance), LTE UEs' channel estimation quality will be slightly degraded from the limited interpolation due to LTE puncturing. However, UEs are generally designed to deal with frequency selective channels and thus will not obtain an average channel estimate over the whole band. Even if some LTE UEs exploit the pilots located in the entire band, the performance degradation can be tolerated within the design margin.

To sum up, though DSR inevitably affects the channel structure of LTE, it appears feasible with a careful design.

\subsection{How to Mitigate the Inter-technology Interference between LTE and GSM}

Since LTE signal and GSM signal are transmitted in the same band, there is GSM adjacent channel interference leakage on LTE UE and vice versa. To mitigate interference from LTE PRBs to GSM, transmit power on the PRBs close to GSM PRBs can be reduced. This also helps mitigate the impact of GSM on LTE since other non-adjacent PRBs can be allocated more power and partially recover the LTE capacity loss because of GSM overlay.

In the next section, we formally formulate the above power adjustment problem and study how to adjust the LTE transmit power accordingly. Before that we note that adjacent channel leakage only captures how much power is transmitted by GSM base station on LTE subcarriers. It is also necessary to consider  the LTE  receiver in-channel selectivity (ICS): LTE receiver FFT will grab some in-band GSM power because of the rectangular windowing effect.
The grabbed in-band GSM power will essentially appear as noise and degrade the SNR of LTE signal, which matters because the GSM PSD is much higher compared to that of LTE. To formalize the above arguments, let us use the continuous time FFT in the receiver for ease of exposition and let $x_g(t)$ be the received GSM signal. Then after FFT,
\begin{align}
X_g (f^*) &= \int_{-T_s/2}^{T_s/2} x_g (t) \exp( j 2\pi f^* t ) d t \notag \\
&=\int_{-\infty}^{\infty} x_g (t) \exp( j 2\pi f^* t )  \textrm{rect}(\frac{t}{T_s}) d t \notag \\
&= \int_{-\infty}^{\infty} X_g (f) T_s \textrm{sinc} ( T_s ( f - f^* ) )  d f, \notag
\end{align}
where $T_s$ is the OFDM symbol time, $\textrm{rect}(t)$ is the rectangle function, i.e., $\textrm{rect}(t) = 1$ if $t\in [-\frac{1}{2}, \frac{1}{2}]$ and zero otherwise, $\textrm{sinc}(t) = \frac{\sin (\pi t)}{\pi t} $, and the last equality is due to the generalized Parseval's theorem. It follows that the ``virtual'' GSM PSD seen by the LTE receiver can be written as
\begin{align}
\tilde{S}_g (f^*) = \int_{-\infty}^{\infty} S_g (f) T_s  \textrm{sinc}^2 ( T_s ( f - f^* ) ) d f,
\label{eq:1}
\end{align}
where $S_g (f)$ is the real received GSM PSD. To gain some insight, let us consider two examples. First, $S_g (f) = N_0/2$ which is the PSD of background white noise. Then
\begin{align}
\tilde{S}_g (f^*) &= N_0/2 \int_{-\infty}^{\infty}   T_s  \textrm{sinc}^2 ( T_s ( f - f^* ) ) d f \notag \\
&= N_0/2 \int_{-\infty}^{\infty} \left(   \frac{1}{\sqrt{T_s}} \textrm{rect} ( \frac{t}{T_s} )   \right)^2    d t \notag \\
&= N_0/2, \notag
\end{align}
which agrees with intuition: white noise still appears ``white''. Now consider another example with arbitrary PSD $S_g (f)$ but $T_s \to \infty$. Then
\begin{align}
\tilde{S}_g (f^*) &=  \int_{-\infty}^{\infty} S_g (f) \lim_{T_s \to \infty} \left(  T_s  \textrm{sinc}^2 ( T_s ( f - f^* ) ) \right) d f \notag \\
&=  \int_{-\infty}^{\infty} S_g (f)        \textrm{sinc} ( T_s ( f - f^* ) )  \delta ( f - f^* )  d f \notag \\
&= \lim_{T_s \to \infty}  S_g (f)        \textrm{sinc} ( T_s ( f - f^* ) ) |_{f=f^*} \notag \\
&= S_g (f^*). \notag
\end{align}
where  we use the fact $\lim_{a\to 0} \frac{1}{a} \textrm{sinc} (\frac{f}{a}) = \delta(f)$ in the second equality.
This implies that if the continuous FFT operation was of infinite duration the ``virtual'' GSM PSD $\tilde{S}_g (f)$ seen by the LTE receiver would be the same as the real GSM PSD $S_g (f)$. Hence the grabbed in-band GSM powers at those LTE OFDM subcarriers  are  zero in this ideal case. In general, this is not true due to the finite windowing effect and  the grabbed in-band GSM power at the $k$-th subcarrier is given by
\begin{align}
N^{(g)}_k = \int_{ (k-\frac{1}{2}) b }^{ (k+\frac{1}{2}) b } \tilde{S}_g (f) d f,
\end{align}
where $b$ denotes the subcarrier bandwidth.

Note that we do not have to worry about the ICS so much from LTE to GSM. In particular, since  LTE PSD is much lower than that of GSM, GSM receiver's adjacent channel selectivity will handle that fairly well.

\section{Transmit Power Allocation in LTE}
\label{sec:transmitpoweralloc}
\begin{figure}
\centering
\includegraphics[width=8cm]{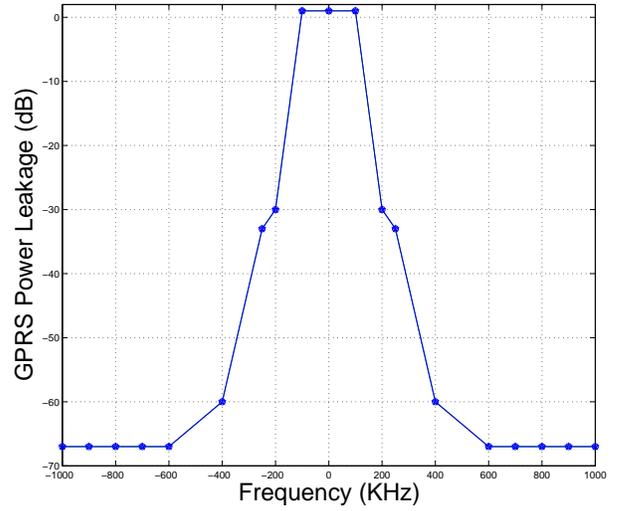}
\caption{Linear interpolation of GPRS adjacent channel interference leakage using data (cf. Table \ref{tab:gprs}) from \cite{ETSIgsm}}
\label{fig:gprs}
\end{figure}

We first model the GSM leakage, which is specified in Table \ref{tab:gprs} and plotted  in Fig. \ref{fig:gprs} \cite{ETSIgsm}. As for  LTE transmit power allocation, the problem formulation involves the PSD of the LTE signal as well as that of the GSM signal. To this end, we first introduce the following notations:


\begin{table*}
\centering
\begin{tabular}{|l|r*{8}{c}|} \hline
$\Delta f$(KHz)              & 100 & 200 & 250 & 400 & 600  & 700 & 800 & 900 & 1000  \\
\hline
ACLR (dB)  & -1  & 30  & 33 & 60  & 67  & 67 & 67  & 67  & 67   \\
\hline
\end{tabular}
\caption{GSM adjacent channel interference leakage specifications \cite{ETSIgsm}}
\label{tab:gprs}
\end{table*}

\begin{itemize}
\item Let $Q$ be the total number of LTE subcarriers, which is assumed to be even.
\item Let $1/T$ be the subcarrier spacing.
\item $\Omega =  \{ \pm 1, \pm 2, ..., \pm Q / 2 \} $ denotes the indices of all the subcarrier frequencies.
\item $S \subset \Omega  $ denotes the set of indices of the subcarrier frequencies that get punctured for GSM overlay.
\item $\Phi = \Omega \backslash S $ denotes the indices of OFDM subcarrier frequencies used for LTE transmissions. Denote by $N = |\Phi|$ the cardinality of $\Phi$.
\item Let $\{1,2,...,N\}$ be the enumeration of $\Phi$. Define the mapping $v_i: \{1,2,...,N\} \mapsto  \Phi$, which maps each $i$ to the corresponding subcarrier frequency index $v_i$.
\item $\{d(n)\}$ denotes the sequence of data symbols and are assumed to be white with $\mathbb{E}[d(n)]=0$, $\mathbb{E}[|d(n)|^2] = 1$.
\end{itemize}
Then the LTE baseband signal $x(t)$ can be written as
$$
x(t) = \sum_{i=1}^N \sqrt{p_i T_s} \sum_k d(kN + i) q_i (t - k T_s),
$$
where $p_i$ is the transmit power on subcarrier $v_i$,  and $q_i (t)$ is defined as
$$
q_i (t) = w(t) \exp ( j 2\pi \frac{v_i}{T} t ),
$$
where $w(t)$ is the windowing function for spectrum shaping.  Then the corresponding PSD is given by
$$
S(f) = \sum_{i=1}^N p_i | W(f - \frac{v_i}{T} ) |^2,
$$
where $W(f)$ is the Fourier transform of $w(t)$.

For practical implementation, the transmitter generates the samples of $x(t)$ at a sufficiently high rate $1/T_c$. These samples are passed through digital-to-analog converter, resulting in $\hat{x} (t)$ given by
$$
\hat{x} (t) = \sum_n x(nT_c) g(t - nT_c),
$$
where $g(t)$ is the interpolation function with Nyquist autocorrelation. Then the corresponding PSD is given by
\begin{align}
\hat{S}(f) &= \frac{1}{T_c^2} |G(f)|^2 \sum_m S( f - \frac{m}{T_c} ) \notag \\
&= \sum_{i=1}^N  p_i \cdot \frac{1}{T_c^2} |G(f)|^2 \sum_m | W(f - \frac{m}{T_c} - \frac{v_i}{T} ) |^2 \notag \\
&= \sum_{i=1}^N  p_i | \hat{W} (f - \frac{v_i}{T} ) |^2, \notag
\end{align}
where $G(f)$ is the Fourier transform of $g(t)$, $\hat{W}(f) = \frac{1}{T_c^2} |G(f)|^2 \sum_m | W(f - \frac{m}{T_c} ) |^2$.

We would like to control the LTE interference leakage to the GSM signal. In particular, the LTE PSD at frequency $f^*$ occupied by the GSM signal should be below some threshold level $C$, i.e.,
$$
S(f^*) = \sum_{i=1}^N W_i (f^*) p_i \leq C.
$$
where
$$
W_i (f^*) = | W(f^* - \frac{v_i}{T} ) |^2.
$$
For discrete-time approximation, one can simply substitute $\hat{S}(f^*)$ for $S(f^*)$ in the above inequality.
As for the LTE signal, its SINR of subcarrier $i$ is
$$
\textrm{SINR}_i = \frac{ p_i }{N_i },
$$
where $N_i$ is the background noise power (possibly including in-band GSM leakage and the grabbed in-band GSM power $N^{(g)}_i$) of the LTE signal in subcarrier $i$.

We are interested in maximizing the LTE spectral efficiency while limiting LTE leakage to GSM signal and meeting the power budget constraint:
\begin{align}
\textrm{maximize}_{p \geq 0} \ \ \  &R(p) = \sum_{i=1}^N \log ( 1 +  \frac{ p_i }{N_i } ) \notag \\
\textrm{subject to} \ \ \   &S(f_j) = \sum_{i=1}^N W_i (f_j) p_i \leq C_j, j=1,...,M, \notag \\
&\sum_{i=1}^N p_i \leq P_{\max},
\label{eq:opt}
\end{align}
where $f_j$ is the $j$-th constrained sample frequency point; $C_j$ is the corresponding threshold value; and $M$ is the total number of sample frequency points. For example, we may take $f_j$ to be the center frequency of each GSM subcarrier. In this example, $M$ equals the number of subcarriers reserved for GSM transmissions. More general constraints are also possible. With convex objective function and convex feasible set, this is a convex programming problem that has a strictly feasible solution and thus can be decoupled across subcarriers. We propose a waterfilling-based search algorithm which exploits the special structure of the above optimization problem. The algorithm can be found in the Appendix.

\begin{figure}
\centering
\includegraphics[width=8cm]{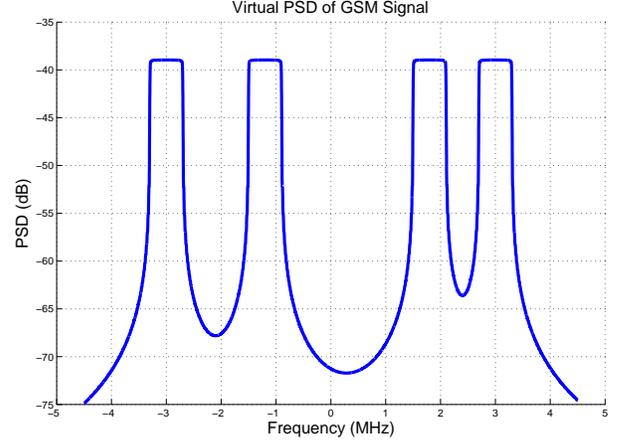}
\caption{Virtual PSD of GSM signal due to the LTE receiver's in-channel selectivity.}
\label{fig:12}
\end{figure}

Next we provide a numerical result to show the structure of the optimal LTE power allocation. 4 portions, each of bandwidth 600KHz, of the $10$MHz LTE channel are punctured. We assume that the GSM PSD is flat in the punctured portions but    $13.56$dB above that of LTE. Fig. \ref{fig:12} shows the virtual PSD of GSM signal ``seen'' by LTE receiver due to ICS. As expected, LTE subcarriers adjacent  to the punctured portions grab more in-band GSM power and thus suffer more from ICS.

\begin{figure}
\centering
\includegraphics[width=8cm]{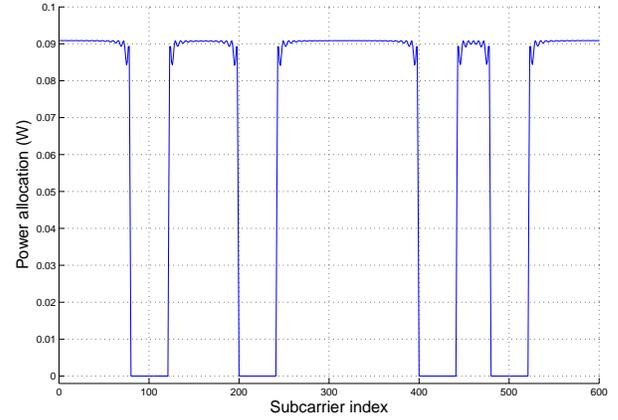}
\caption{Optimal LTE power allocation: $T_s/T = 1200/1024$ with $T = \frac{1}{1200}\times 10^{-3}$, total transmit power $P_{\textrm{max}} = 40W$, noise power per subcarrier is $1/(10*T_s*192000)W$. (Thus, the ratio of total transmit power to total noise power is $10$dB.)  }
\label{fig:17}
\end{figure}

\begin{figure}
\centering
\includegraphics[width=8cm]{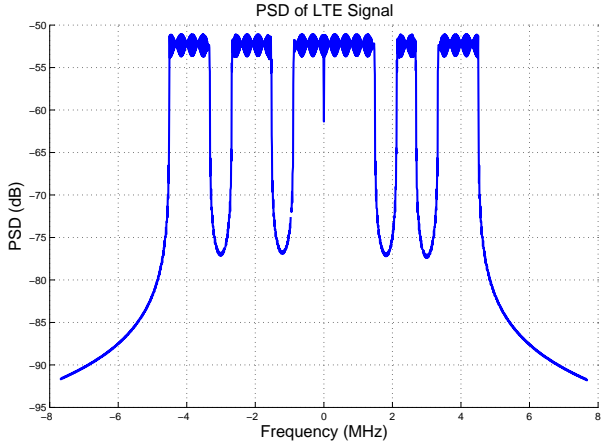}
\caption{PSD of the LTE signal: LTE leakage to the 4 punctured spectrum portions reserved for GSM is limited below $-15$dB.}
\label{fig:11}
\end{figure}

The optimal power allocation solution to the problem in (\ref{eq:opt}) is plotted in Fig. \ref{fig:17}, while the PSD of the punctured LTE signal is shown in Fig.\ref{fig:11}. The optimal power allocation shown in Fig. \ref{fig:17} is \textit{almost} uniform power allocation across the spectrum except zero power in the punctured spectrum and nearly zero power in the nearest two subcarriers on each side of the 4 punctured portions of the  spectrum. It can be seen from Fig. \ref{fig:11}  that LTE leakage to GSM is limited below $-15$dB and thus the GSM channels are well protected. The design insight here is thus to  reserve 2 subcarriers as the guard band between GSM and LTE at each side of every punctured portion of the spectrum. This would protect GSM signals from LTE leakage. Note that in our optimization formulation we only protect GSM signals from LTE leakage. Of course, this is only one aspect of the design; more subcarriers may need to be reserved in real systems for other concerns.

\begin{figure}
\centering
\includegraphics[width=8cm]{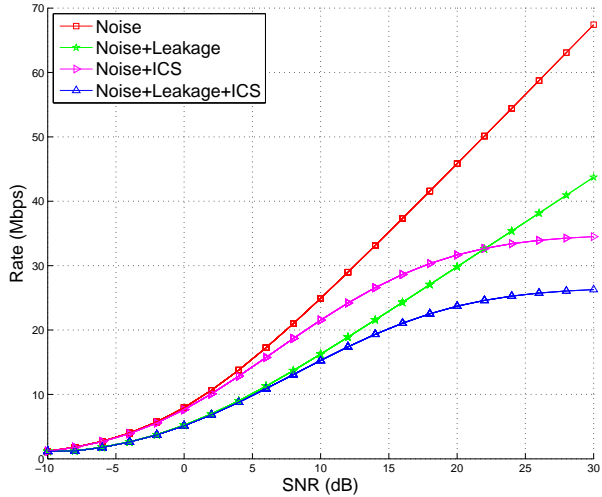}
\caption{Impact of GSM power leakage and LTE receiver's ICS: SNR is defined as the ratio of total LTE received power to total noise power in 9MHz LTE channel.}
\label{fig:13}
\end{figure}

Fig. \ref{fig:13} shows the impact of GSM power leakage and LTE receiver's ICS. It can be seen that GSM power leakage  significantly reduces the LTE rate. Besides GSM power leakage, LTE receiver's ICS further reduces the rate, particularly in the high SNR regime, i.e., SNR$>$15dB. In real LTE networks, SNR is typically much lower than $15$dB, which implies that  from a system wide perspective GSM power leakage is a more dominant issue compared to LTE receiver's ICS. Note that the Noise+Leakage  and Noise+ICS  curves cross at high SNR, which arises from the following facts: In the low SNR regime the limiting factor is the bad SINR of adjacent PRBs due to GSM power leakage, while in the high SNR regime ICS dominates because a $-72-(-39) = -33$dB floor (cf. Fig. \ref{fig:12}) exists throughout the LTE channel, which eventually limits the achievable rate.

Before ending this section, we stress that the above power optimization is formulated from the link-level perspective. Compared to a system-wide perspective, the link-level power control problem is relatively simple; better performance can be obtained with system-level power optimization. Indeed, semi-static power allocation over different PRBs is supported by LTE for the purpose of inter-cell interference coordination (ICIC). The proposal in this paper can be combined with such heterogeneous network concepts. In addition, the focus here is on a commonly used rate metric to demonstrate the benefits on spectrum sharing on throughput. Other metrics like delay may become relevant in system-level study, where multiple queues exist and scheduling plays an important role in determining the delay performance. Though desired delay performance is normally achieved by designing appropriate scheduling algorithms (see e.g. \cite{parekh1993generalized, fattah2002overview, liu2006cross}), delay can be incorporated into the power optimization by following e.g. \cite{li2009optimal}. Roughly speaking, one can come up with an (possibly approximate) expression for the mean delay of each queue as a function of the power allocation. Then for each link one can set a maximum delay constraint. The difficulty of the delay-incorporated power optimization will be largely determined by the delay constraints (which determine the feasible set of power allocation) and deserves future study.

\section{LTE Enhancements to Recover Capacity Loss}
\label{sec:lte}

Due to GSM overlay, there is a capacity  loss in the LTE system. This loss stems from the reduced number of PRBs that can be used for LTE UEs as well as from the GSM interference due to leakage and LTE receiver's ICS. In this section, we investigate ways for LTE to recover the capacity loss.

\subsection{Fractional Frequency Reuse}

Since adjacent GSM cells/sectors use orthogonal carriers, FFR can be applied between LTE and GSM to mitigate the capacity loss of LTE. In particular, LTE in a given cell can utilize  GSM carriers in the neighboring cells with low transmit power. Low power LTE transmissions targeting good geometry UEs can provide significant spectral efficiency. Nevertheless, we need to determine how low the power should be so as not to affect  GSM transmissions in the neighboring cells.

\begin{figure}
\centering
\includegraphics[width=8cm]{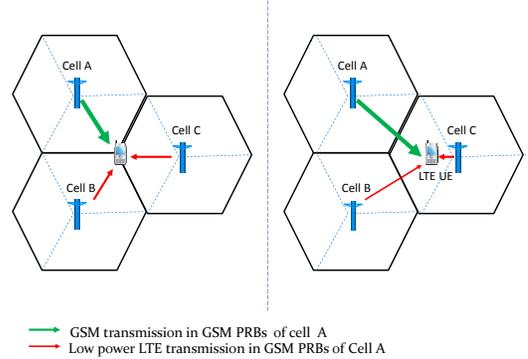}
\caption{Fractional frequency reuse of GSM PRBs in neighboring cells}
\label{fig:lowPower}
\end{figure}
We next roughly calculate this low power. Consider the worst scenario as shown in the left plot in Fig. \ref{fig:lowPower}. Let $P_g$ and $P_s$ denote the GSM transmit power and the LTE low power (in a 200KHz GSM channel),
respectively. To guarantee reliable GSM transmission, we require
\begin{align}
\frac{P_g H}{2P_s H + N_0} \geq \gamma \Rightarrow P_s  \leq \frac{P_g }{2\gamma} - \frac{N_0}{2H}
\label{eq:gprs}
\end{align}
where $\gamma$ is the required SINR threshold for reliable GSM transmission, e.g. 10dB, $P_g$ is the GSM transmit power, $H$ is the mean channel gain (i.e., only path loss is considered) which is the same for all the $3$ links in the left plot in Fig. \ref{fig:lowPower}, $N_0$ is the noise power per 200KHz GSM channel.
Based on (\ref{eq:gprs}), we can at least get a rough estimate about the low power level that could be used in FFR.

The right hand side plot in Fig. \ref{fig:lowPower} shows that the low power LTE transmission in cell C suffers from intercell interference caused by cochannel GSM transmission in cell A and low power LTE transmission in cell B (since cell B also reuses the GSM PRBs of cell A).
However, compared to the interference caused by the neighboring cochannel GSM transmission, interference generated by the neighboring low power LTE transmission is negligible since it uses a much lower transmit power.

\subsection{Intelligent Scheduling}

There are two types of ``special'' PRBs in LTE that are overlaid by GSM and enhanced with FFR. The first type includes those PRBs adjacent to GSM PRBs. Due to the GSM interference leakage, the maximum average SINR of LTE UE in these PRBs is limited to $11$dB, as pointed out before. In LTE network, the SINR of $50\%$ of UEs is less than 1dB because of out-of-cell interference. As a result, from system point of view, scheduling cell edge UEs (bottom 50\%) in PRBs adjacent to GSM PRBs will result in nearly no loss of spectral efficiency in LTE PRBs.

The second type refers to those fractional reused PRBs with low transmit power. Since these PRBs are allocated with low power and suffer from strong GSM interference from the neighboring cell, we ought to schedule UEs of good geometry\footnote{Though referred to as good \textit{geometry} UE, a better quantification may be based on SINR, e.g., UEs with top $x\%$ ($x\in[5,15]$) SINR in a network. Typically, the interference experienced  by these UEs is small and may be ignored for discussion purpose.} in these reused PRBs to maximize the spectral efficiency.

The above discussion seems to imply that we need to revise the existing scheduling algorithms. However, the current proportional fair (PF)  scheduling (see, e.g., \cite{Tse2005Fundamentals, Kushner2004Convergence, Stolyar2005on}) exploits the multiuser diversity and we find that a frequency-selective PF scheduler performs pretty well in LTE with GSM overlay, as demonstrated by simulation results in Section \ref{sec:sim}. 

Specifically, the generalized frequency-time PF scheduling works as follows. In time slot $n$, the LTE eNodeB computes the achievable instantaneous rates $R_k^m(n)$ for UE $k$ at time $n$ in PRB $m$. Then the scheduler allocates the PRB $m$ to the UE $k_m^\star$ with the largest
\begin{align}
P_k^m (n) = \frac{ R_k^m(n) }{ T_k (n) } \notag
\end{align}
among all the associated UEs. Here $T_k (n)$ denotes the average throughput of UE $k$ up to time $n$ and is updated according to the following weighted low-pass filer:
\begin{align}
T_k (n+1) = (1 - \frac{1}{N_T}) T_k (n) + \frac{1}{N_T} \sum_{m} \delta(k-k_m^\star) R_k^m(n), \notag
\end{align}
where $\delta(x)=1$ if $x=0$ and zero otherwise, and $N_T$ denotes the response time of the low-pass filer.


As a high level understanding on the good performance of the frequency-selective PF scheduler, let us compare good geometry UEs and other UEs in the reused PRBs. Note that the intercell interference does not matter much for good geometry UEs and thus is ignored in the following high level discussion. Thus, the instantaneous achievable rate of good geometry UEs in the reused PRBs would roughly be given by
$$
\log(1 + \frac{1}{\Gamma} \frac{P_s H}{N_0} )
$$
where $\Gamma$ is the Shannon gap and $N_0$ is the noise power per PRB (180KHz) channel. In contrast, the instantaneous achievable rate of other UEs in the reused PRBs would be given by
$$
\log(1 + \frac{1}{\Gamma} \frac{P_s \hat{H}}{I_g + I_l + N_0} )
$$
where $H \gg \hat{H}$. Thus, it appears that PF scheduler would schedule good geometry UEs in the reused PRBs more often than other UEs. This is also demonstrated through simulation in Section~\ref{sec:sim}.

To sum up, though one may consider designing novel scheduling algorithms to achieve better performance, PF scheduling is sufficient for initial evaluation of the proposed DSR.


\subsection{Partial PRB Usage}

\begin{figure}
\centering
\includegraphics[width=8cm]{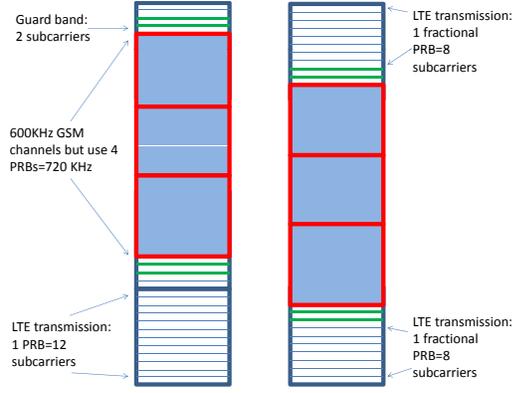}
\caption{Partial PRB usage: On the left plot, only one PRB is available for LTE transmission. On the right plot, two fractional PRBs (or equivalently, 4/3 PRBs) can be used for LTE transmission. }
\label{fig:fractionalPRB}
\end{figure}

Recall that GSM center frequency must be an integer multiple of 200KHz while the bandwidth of each PRB, the basic LTE scheduling unit, is 180KHz. Hence, GSM channel edges may not be always aligned with LTE PRB edges. For example, suppose we puncture the LTE spectrum for a 600KHz GSM band. Then 4 PRBs would have to be reserved, leading to 120KHz unutilized spectrum as it is only a fraction of one PRB and thus can not be scheduled. This simple numerical example is illustrated in Fig. \ref{fig:fractionalPRB}, which further takes guard band into account. Fig. \ref{fig:fractionalPRB} shows that better spectrum utilization can be achieved if the system regains access to the fractional PRBs resulted from GSM overlay. 

Motivated by the above observation, we propose the concept of partial PRB usage. One possible implementation of partial PRB usage may work as follows: In the MAC layer LTE scheduler treats a fractional PRB as a normal PRB and allocates the whole PRB to some UE based on any implemented scheduling algorithm; in the PHY layer, LTE eNodeB allocates zero power on the subcarriers used by GSM channels as well as the guard band. Using this implementation, the impact of partial PRB usage is equivalent to puncturing the codewords of those UEs using partial PRBs. So we might need to slightly decrease the coding rate of those UEs in the rate matching step during downlink transport channel processing to maintain similar bit error rate (BER) performance.

To sum up, partial PRB usage essentially enables overlaying GSM channels on LTE channel on subcarrier basis rather than on PRB basis. Regaining access to the fractional PRBs resulted from GSM overlay can provide better spectrum utilization. In addition,  subcarrier-based overlay enjoys higher flexibility in selecting the positions of the punctured spectrum in LTE channel.

\section{Simulation Results}
\label{sec:sim}

In this section we provide simulation results to evaluate the impact of GSM overlay on LTE spectral efficiency. The network layout used in simulation consists of regular hexagonal cells, each of which is composed of 3 sectors.
The center sector is treated as the typical one, while the remaining sectors act as interfering sources. The radiation pattern used for the directional antennas is as follows:
\begin{align}
G(\phi) = \frac{1}{2} ( 1 + \cos \phi ) \cdot \frac{ \sin ( 0.72\pi \sin \phi ) }{ 0.72\pi \sin \phi }.
\label{eq:2}
\end{align}
The GSM transmit power is measured by its PSD offset to the PSD of LTE signal; it is set to be $13.56$dB in the simulations.

The system simulation conducted is running at the granularity of transmission time interval (TTI) of 1ms, based on which scheduling decisions are made. The statistics are collected from 3 drops of the UEs.
The main simulation parameters used are summarized in Table \ref{tab:sys:para} unless otherwise specified.
\begin{table}
\centering
\begin{tabular}{|l||r|} \hline
$\#$ of hexagonal cells & $6 \times 6$  \\ \hline
Length of cell edge  & $500$m  \\ \hline
$\#$ of sectors per cell & $3$  \\ \hline
$\#$ of UEs per sector   & 24  \\ \hline \hline
Max TX power & $40$W \\ \hline
Noise PSD & $-174$dBm \\ \hline
Wavelength $\lambda$ & 0.375m \\ \hline
Path loss exponent $\alpha$ & 3 \\ \hline
Shadowing deviation $\sigma^2$ & 36 \\ \hline \hline
LTE transmission bandwidth & $10$MHz \\ \hline
Subcarrier bandwidth $B$ & $15$KHz \\ \hline
$\#$ of PRBs & $50$ \\ \hline \hline
GSM power offset & $13.56$dB \\ \hline
GSM carrier bandwidth & $200$KHz \\ \hline
GSM SINR threshold $\gamma$ & $10$dB \\ \hline \hline
Gap $\Gamma$ & $3$dB \\ \hline
\end{tabular}
\caption{Simulation parameters}
\label{tab:sys:para}
\end{table}

\begin{figure}
\centering
\includegraphics[width=8.8cm]{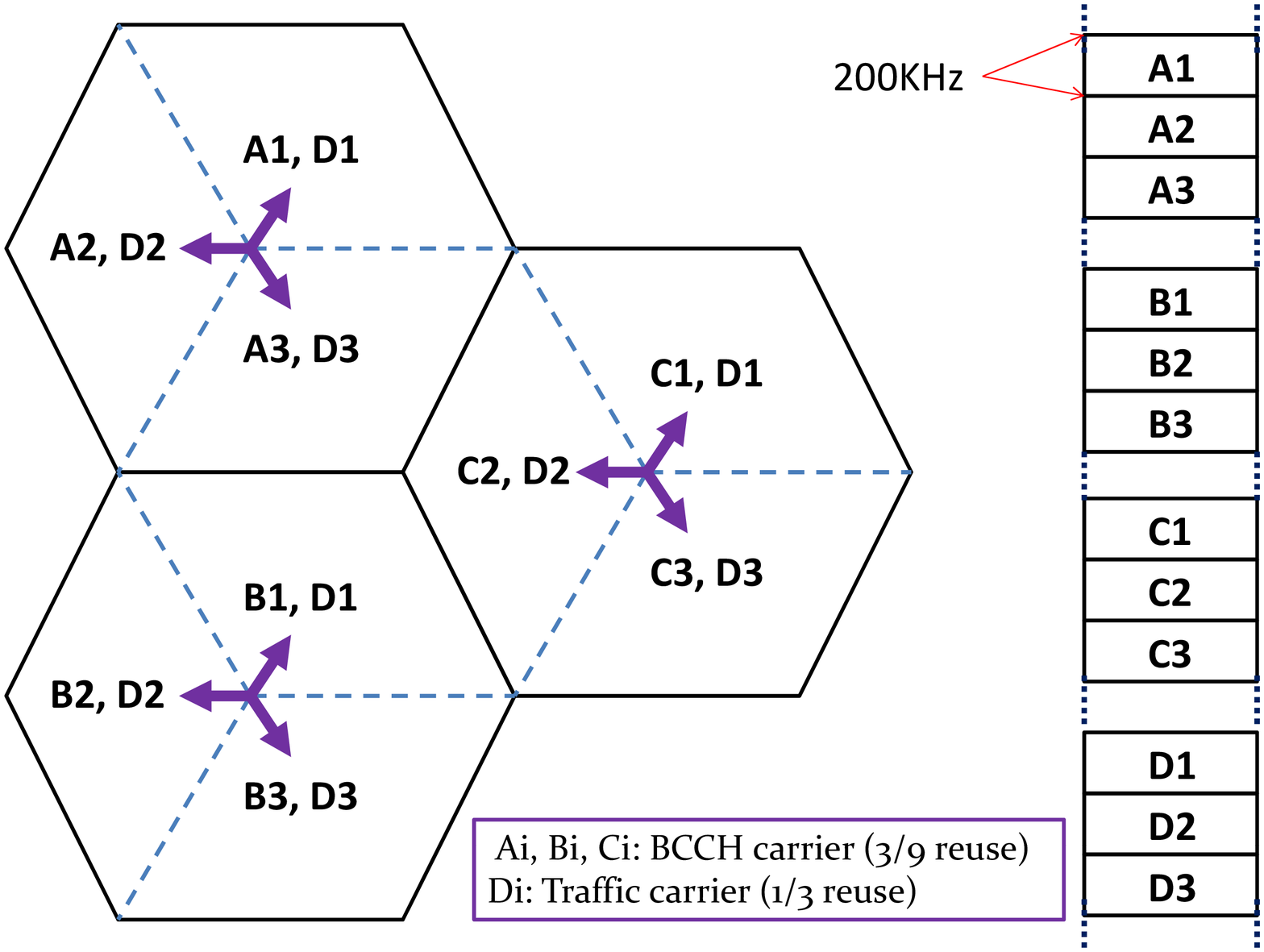}
\caption{LTE puncturing and GSM frequency planning: 10MHz LTE channel with a 2.4MHz punctured portion reserved for GSM with 3/9 reuse for BCCH carriers and 1/3 reuse for traffic carrier.}
\label{fig:reuse}
\end{figure}

The specific LTE puncturing and GSM frequency planning used in the simulation are shown in Fig.\ref{fig:reuse}. In particular, there is one BCCH carrier and one traffic carrier in each sector for GSM transmission. The reuse factor is 3/9 for BCCH carriers and is 1/3 for traffic carrier. Note that with this particular frequency planning, a sector, say the one with carrier A3 and D3 in Fig.\ref{fig:reuse} can reuse all the carriers denoted by Bi,Ci, $i=1,2,3$. Partial PRB usage is not needed with this particular GSM frequency planning. For example, the sector labeled by ($C_2,D_2$) in Fig.\ref{fig:reuse} can fully reuse the originally unused $2\times 720$KHz spectrum reserved for GSM carriers $A_i$ and $B_i$.

\begin{figure}
\centering
\includegraphics[width=8.8cm]{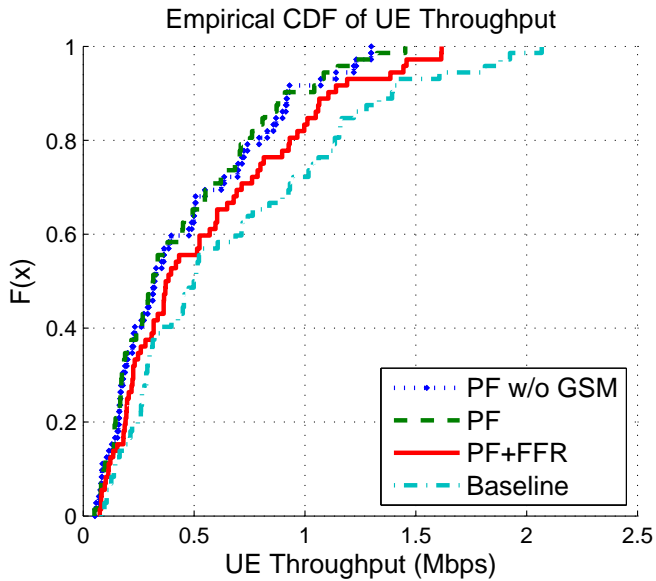}
\caption{Empirical CDF of UE throughput.}
\label{fig:pfCDF}
\end{figure}

We first would like to show that the LTE capacity loss is less than $2.4/9 \approx 26\%$, since a portion of 2.4MHz in the 9MHz \textit{occupied} LTE bandwidth is reserved for GSM. To this end, the following 4 scenarios are studied and compared in the simulation:
\begin{itemize}
\item Baseline: This corresponds to the normal 10MHz LTE deployment without puncturing.
\item PF w/o GSM: This corresponds to the 10MHz LTE deployment with 2.4MHz puncturing but without GSM overlay. This scenario is considered purely for comparison.
\item PF: This corresponds to the 10MHz LTE deployment with 2.4MHz puncturing and GSM overlay.
\item PF+FFR: This corresponds to the 10MHz LTE deployment with 2.4MHz puncturing and GSM overlay and FFR enhancement.
\end{itemize}
The LTE rate statistics are summarized in Table \ref{tab:rate} and the associated empirical CDFs of UE throughput are shown in Fig. \ref{fig:pfCDF}. We make the following remarks:
\begin{itemize}
\item Compared to PF w/o GSM, we can see only very minor performance degradation exists in PF for mean rate of all the UEs and no performance degradation for mean rates of both top and bottom 5\% of UEs, as shown in Table \ref{tab:rate}. In addition, they have very similar CDFs of UE throughput.
\item When PF is jointly applied with FFR, the loss of the mean rate of all the UEs is less than $20\%$. 
This shows that the loss of LTE capacity caused by GSM overlay can be significantly reduced with the proposed enhancements.
\item Compared to PF, there is  a rate loss for bottom-5\% UEs in PF+FFR. This shows that while FFR improves the mean rate of all the UEs by reusing the spectrum more, it does lead to poorer SINR and thus degraded throughput to cell edge users.
\item The CDFs of UE throughput for PF w/o GSM, PF, and PF+FFR have very minor difference for the bottom 50\% UEs, i.e., the range with $F(x) \leq 0.5$ in Fig. \ref{fig:pfCDF}.
\end{itemize}

\begin{figure}
\centering
\includegraphics[width=8cm]{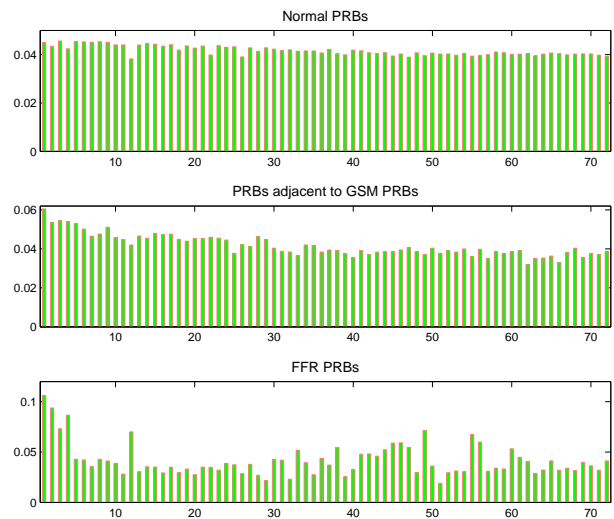}
\caption{PRB scheduling statistics.}
\label{fig:scheduling}
\end{figure}

We next show the scheduling statistics of normal PRBs, PRBs adjacent to GSM PRBs, and FFR PRBs in   Fig.\ref{fig:scheduling}, where users are ranked according to their rates. We can see that the allocation probability of a normal PRB is almost uniform among the UEs. This is consistent with the fact that PF scheduling maintains fairness in the long run. In contrast, the allocation probability of a PRB adjacent to GSM PRBs is not uniform among the UEs. When it comes to FFR PRBs, as expected, good geometry UEs have relatively higher probabilities to be allocated with FFR PRBs.

\begin{table*}
\centering
\begin{tabular}{|l|l||r||r|r|r|}
\hline
Mean rate &    & PF w/o GSM  & PF & PF+FFR &  Baseline  \\ \hline
\multirow{2}{*}{All the UEs} & Mbps  & 0.44999  & 0.44189  & 0.54237   & 0.66991 \\ \cline{2-6}
                             & $\%$  & 67.17 & 65.96   & 80.96 & 100   \\
 \hline
  \multirow{2}{*}{Top-$5\%$ UEs} & Mbps  & 1.2381 & 1.2471 & 1.5041 & 1.8588  \\ \cline{2-6}
                                    & $\%$  & 66.61 & 67.09 & 80.92 & 100  \\
 \hline
 \multirow{2}{*}{Bottom-$5\%$ UEs} & Mbps  & 0.06477 & 0.06620 & 0.05295 & 0.09763  \\ \cline{2-6}
                                   & $\%$  & 66.35 & 67.79 & 54.24 & 100  \\
\hline
\end{tabular}
\caption{LTE Enhancements to recover capacity loss.}
\label{tab:rate}
\end{table*}

\section{Discussion on Uplink Design}
\label{sec:uplink}

Though in this paper we focus on describing and evaluating the proposed DSR approach in the downlink, we discuss in this section on the various issues involved in the uplink design. 

\subsection{Reserving PRBs}

Compared to the downlink, the LTE uplink channel structure is simpler \cite{3gppMobility}. This makes the selection of the reserved PRBs out of the LTE uplink channel easier. 

 \begin{figure}
 \centering
 \includegraphics[width=8cm]{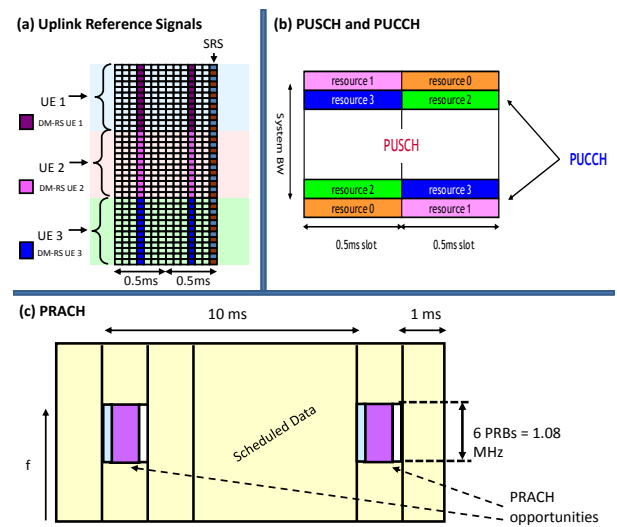}
 \caption{LTE uplink channel structure}
 \label{fig:uplink}
 \end{figure}

\subsubsection{Impact on reference signals} 

Two types of reference signals exist in the uplink: Demodulation Reference Signals (DM-RS) and Sounding Reference Signals (SRS). LTE eNodeB relies on DM-RS for coherent demodulation. As shown in Fig. \ref{fig:uplink}(a), each UE's DM-RS is time multiplexed with the corresponding uplink data and occupies the same bandwidth as the data. In LTE, UE transmissions are localized by Single-carrier Frequency-Division Multiple Access (SC-FDMA); thus, DM-RS would not be affected by LTE puncturing as the reserved PRBs will not be scheduled for any UE and thus the corresponding DM-RS is suppressed.

The SRS is used by LTE eNodeB to estimate the uplink channel quality, based on which the UEs are scheduled across the frequency domain. Unlike DM-RS occupying the same bandwidth as each UE's data, SRS of each UE stretches out the allocated bandwidth for wider channel quality estimation. The SRS transmission mode can be either wideband or narrowband. As a single SRS transmission is used in wideband mode, LTE puncturing may restrict the frequency range of channel estimation. From this perspective, we suggest using the narrowband option, in which SRS transmission is split into a series of narrowband transmissions. Thus, the narrowband option is flexible enough to cover the frequency range of interest for channel estimation.

\subsubsection{Impact on uplink physical channels} 

Three types of physical channels exist in the uplink: Physical Uplink Shared Channel (PUSCH), Physical Uplink Control Channel (PUCCH) and Physical Random Access Channel (PRACH). PUSCH carries UE data as well as certain associated control information. For similar reason as in the DM-RS case, PUSCH would not be affected by LTE puncturing.

PUCCH carries control signaling such as ACK/NAK, channel quality indicators (CQI) and scheduling requests. Unlike its downlink counterpart PDCCH stretching over the whole transmission bandwidth, Fig. \ref{fig:uplink}(b) shows that PUCCH is sent on the band edges of the system bandwidth unless the over-provisioned Physical Uplink Control Channel (PUCCH) (i.e., moving PUCCH away from the LTE channel edges) is supported. So we can avoid affecting PUCCH by not reserving the PRBs used by PUCCH.

PRACH carries the random access preamble sent by a UE for asynchronous network access. As shown in Fig. \ref{fig:uplink}(c), it occupies 1.08 MHz  contiguous bandwidth (i.e. 6 contiguous PRBs). The specific PRACH resources are assigned by LTE eNodeB within PUSCH region. Thus, we can reserve the PRBs for GSM without overlapping with the resources used by PRACH.

\subsection{Mitigating Mutual Interference between LTE and GSM}

As in the downlink, there exists mutual interference between LTE and GSM. 

\subsubsection{GSM uplink transmission on LTE reception}

Due to adjacent channel leakage and ICS, GSM terminals will cause interference to LTE UEs. The interference can be especially strong from those GSM terminals close to LTE eNodeB, which can limit the LTE eNodeB's ability to detect weak LTE uplink signals. This leads a near-far problem between GSM and LTE, analogous to that in a CDMA system. This problem can be partially solved by GSM power control: GSM terminals close to LTE eNodeB reduce their transmit power.

Ideally, it is expected that all the GSM received signal strengths are roughly at the same level. This may  be hard to achieve due to the crude GSM power control. Thus, LTE closed loop power control is also crucial to overcome GSM interference. Specifically, the LTE eNodeB measures the received SINR, and compares it to the target value. Then UE-specific power offsets can be sent to UEs via Radio Resource Control (RRC) signaling. These transmit power control commands can be used by LTE UEs to correct open-loop errors or to allow proprietary methods to create a power profile. Further aperiodic fast power control is also possible by additionally allowing a dynamic adjustment of the UE transmit PSD with 1 or 2 bit power control commands.

Note that the UEs scheduled on those PRBs adjacent to the GSM PRBs are the main victims suffering from GSM interference; thus with power control they use higher transmit power. In order to mitigate the out-of-cell interference from these higher power transmissions, it will be helpful to schedule LTE UEs close to the eNodeB in PRBs adjacent to GSM PRBs.

\subsubsection{LTE uplink transmission on GSM reception}

Adjacent channel interference impact of LTE uplink on GSM reception will be similar to that in downlink. Considering the facts that GSM terminal close to the BS may blast significant amount of power (since the power control in GSM is crude), and that the PSD of LTE UEs are generally lower (since LTE UE uplink transmissions are wideband while GSM transmissions are narrowband), LTE interference will be less of an issue to GSM reception.

\section{Conclusions}
\label{sec:conclusion}

In this paper we propose a novel solution to provide GSM connectivity within an LTE carrier through an efficient, dynamic overlay by reserving a few physical resource blocks for GSM. With this approach, operators can migrate their 2G spectrum to LTE while still providing reduced capacity GSM connectivity to their low data rate M2M customers. Furthermore, spectrum can be dynamically shared between LTE and GSM. The system design and feasibility of DSR are carefully detailed. Several enhancements including inter-technology ICIC, intelligent scheduling and partial PRB usage are proposed to recover LTE capacity loss due to the GSM overlay. Our study shows that the loss of LTE capacity caused by GSM overlay can be significantly reduced with the proposed enhancements. Though the focus of this paper is about GSM networks, the same ideas can be applied for other narrow band technology overlays over LTE.

\appendix[Optimizing LTE Transmit Power Allocation]

The general optimal power allocation problem is convex with linear constraints. The feasible set is an $N$-dimensional polytope and the optimal point lies on the face defined by either hyperplane $\sum_{i=1}^N W_i (f_j) p_i = C_j$ or hyperplane $\sum_{i=1}^N p_i = P$.  This suggests a simple algorithm to compute the optimal power allocation as follows.

\begin{enumerate}
\item Initialize $\Psi = \{0,1,...,M\}$.
\item Let $$
p^0_i = \left( \frac{1}{\mu} - N_i  \right)^+, \quad i=1,...,N,
$$
where $\mu$ is chosen such that
$$
\sum_{i=1}^N \left( \frac{1}{\mu} - N_i  \right)^+ = P_{\max}.
$$
Compute the associated spectral efficiency as
$$
R_0 = \sum_{i=1}^N \log ( 1 +  \frac{  p^0_i }{N_i } ).
$$
\item For $j=1,...,M$,
\begin{itemize}
\item Check if $\sum_{i=1}^N W_i (f_j) p^0_i \leq C$.
\item If yes, $\Psi = \Psi \backslash \{j\}$.
\item Otherwise,
$$
p^j_i = \left( \frac{1}{W_i (f_j) \lambda_j} - N_i  \right)^+, \quad i=1,...,N,
$$
where $\lambda_j$ is chosen such that
$$
\sum_{i=1}^N \left( \frac{1}{W_i (f_j) \lambda_j} - N_i  \right)^+ = C_j.
$$
Compute the associated spectral efficiency as
$$
R_j = \sum_{i=1}^N \log ( 1 +  \frac{  p^j_i }{N_i } ).
$$
\end{itemize}
\item Find $j^\star = \arg \max_{j \in \Psi} R_j $. Output the optimal solution:
$
p^\star = p^{j^\star}
$
\end{enumerate}

\textbf{Remark:} The main computation burden associated is to calculate waterfilling solutions, which can be computed efficiently using, e.g., the framework proposed in \cite{Palomar2005Practical}. The number of times to invoke waterfilling algorithm lies in $[1,M+1]$. The above algorithm is viable when $M$ is not large since the waterfilling solution can be obtained very efficiently. For the problem of GSM overlay on LTE, we puncture 9MHz LTE channel to have 12 GSM channels with bandwidth 200KHz each. Every three of the GSM channels are in the same group. If we could protect the GSM transmission of the leftmost and rightmost GSM channels in one group, the GSM signal of the middle carrier in this group would automatically get protected. So we only have $M=8$ GSM SINR constraints. This enables the above search algorithm suitable in this paper.

\bibliographystyle{IEEEtran}
\bibliography{IEEEabrv,Reference}

\begin{thebibliography}{10}
\providecommand{\url}[1]{#1}
\csname url@samestyle\endcsname
\providecommand{\newblock}{\relax}
\providecommand{\bibinfo}[2]{#2}
\providecommand{\BIBentrySTDinterwordspacing}{\spaceskip=0pt\relax}
\providecommand{\BIBentryALTinterwordstretchfactor}{4}
\providecommand{\BIBentryALTinterwordspacing}{\spaceskip=\fontdimen2\font plus
\BIBentryALTinterwordstretchfactor\fontdimen3\font minus
  \fontdimen4\font\relax}
\providecommand{\BIBforeignlanguage}[2]{{%
\expandafter\ifx\csname l@#1\endcsname\relax
\typeout{** WARNING: IEEEtran.bst: No hyphenation pattern has been}%
\typeout{** loaded for the language `#1'. Using the pattern for}%
\typeout{** the default language instead.}%
\else
\language=\csname l@#1\endcsname
\fi
#2}}
\providecommand{\BIBdecl}{\relax}
\BIBdecl

\bibitem{cisco2011cisco}
Cisco, ``Cisco visual networking index: Global mobile data traffic forecast
  update, 2011-2016,'' \emph{white paper}, February 2012.

\bibitem{website:3gppLTE}
\BIBentryALTinterwordspacing
3GPP, ``{LTE} official website,'' 2012. [Online]. Available:
  \url{http://www.3gpp.org/LTE}
\BIBentrySTDinterwordspacing

\bibitem{Astely2009LTE}
D.~Astely, E.~Dahlman, A.~Furuskar, Y.~Jading, M.~Lindstrom, and S.~Parkvall,
  ``{LTE}: the evolution of mobile broadband,'' \emph{IEEE Communications
  Magazine}, vol.~47, no.~4, pp. 44--51, April 2009.

\bibitem{Ghosh2010Fundamentals}
A.~Ghosh, J.~Zhang, R.~Muhamed, and J.~G. Andrews, \emph{Fundamentals of
  LTE}.\hskip 1em plus 0.5em minus 0.4em\relax Prentice Hall, 2010.

\bibitem{Vodafone2010M2M}
Vodafone, ``Global machine to machine communication,'' \emph{white paper},
  2010. Available at goo.gl/4V8az.

\bibitem{Ericsson2011M2M}
Ericsson, ``Device connectivity unlocks value,'' \emph{white paper}, January
  2011. Available at goo.gl/alou6.

\bibitem{mehrotra1997gsm}
A.~Mehrotra, \emph{GSM System Engineering}.\hskip 1em plus 0.5em minus
  0.4em\relax Artech House Inc., 1997.

\bibitem{GSMModules-ABI}
{ABI Research}, ``Cellular {M2M} connectivity services,'' Tech. Rep., 2012.

\bibitem{lin2013dynamic}
X.~Lin and H.~Viswanathan, ``Dynamic spectrum refarming of {GSM} spectrum for
  {LTE} small cells,'' \emph{arXiv preprint arXiv:1305.2999}, 2013.

\bibitem{ETSIgsm}
ETSI, ``{Digital cellular telecommunications system (Phase 2+); Radio
  transmission and reception (GSM 05.05 version 5.12.0 Release 1996)},''
  \emph{ETSI TS 100 910 V5.12.0}, February 2001.

\bibitem{Iamura2010CA}
M.~Iwamura, K.~Etemad, M.-H. Fong, R.~Nory, and R.~Love, ``Carrier aggregation
  framework in {3GPP} {LTE-Advanced} [{WiMAX/LTE} update],'' \emph{IEEE
  Communications Magazine}, vol.~48, no.~8, pp. 60--67, August 2010.

\bibitem{Shen2012over}
Z.~Shen, A.~Papasakellariou, J.~Montojo, D.~Gerstenberger, and F.~Xu,
  ``Overview of {3GPP} {LTE-Advanced} carrier aggregation for {4G} wireless
  communications,'' \emph{IEEE Communications Magazine}, vol.~50, no.~2, pp.
  122--130, February 2012.

\bibitem{lin2012modeling}
X.~Lin, J.~G. Andrews, and A.~Ghosh, ``Modeling, analysis and design for
  carrier aggregation in heterogeneous cellular networks,'' \emph{arXiv
  preprint arXiv:1211.4041}, 2012.

\bibitem{Boudreau2009Interference}
G.~Boudreau, J.~Panicker, N.~Guo, R.~Chang, N.~Wang, and S.~Vrzic,
  ``Interference coordination and cancellation for {4G} networks,'' \emph{IEEE
  Communications Magazine}, vol.~47, no.~4, pp. 74--81, April 2009.

\bibitem{Stolyar2008Self}
A.~L. Stolyar and H.~Viswanathan, ``Self-organizing dynamic fractional
  frequency reuse in {OFDMA} systems,'' in \emph{Proceedings of IEEE Infocom},
  April 2008, pp. 691--699.

\bibitem{Ali2009ffr}
S.~Ali and V.~Leung, ``Dynamic frequency allocation in fractional frequency
  reused {OFDMA} networks,'' \emph{IEEE Transactions on Wireless
  Communications}, vol.~8, no.~8, pp. 4286--4295, August 2009.

\bibitem{3gppMobility}
3GPP, ``Evolved universal terrestrial radio access ({E-UTRA}); physical
  channels and modulation,'' \emph{3GPP TS 36.211 V10.5.0}, June 2012.

\bibitem{parekh1993generalized}
A.~K. Parekh and R.~G. Gallager, ``A generalized processor sharing approach to
  flow control in integrated services networks: {The} single-node case,''
  \emph{IEEE/ACM Transactions on Networking}, vol.~1, no.~3, pp. 344--357, June
  1993.

\bibitem{fattah2002overview}
H.~Fattah and C.~Leung, ``An overview of scheduling algorithms in wireless
  multimedia networks,'' \emph{IEEE Wireless Communications}, vol.~9, no.~5,
  pp. 76--83, October 2002.

\bibitem{liu2006cross}
Q.~Liu, X.~Wang, and G.~B. Giannakis, ``A cross-layer scheduling algorithm with
  qos support in wireless networks,'' \emph{IEEE Transactions on Vehicular
  Technology}, vol.~55, no.~3, pp. 839--847, May 2006.

\bibitem{li2009optimal}
Y.~Li, M.~Chiang, A.~R. Calderbank, and S.~N. Diggavi, ``Optimal
  rate-reliability-delay tradeoff in networks with composite links,''
  \emph{IEEE Transactions on Communications}, vol.~57, no.~5, pp. 1390--1401,
  May 2009.

\bibitem{Tse2005Fundamentals}
D.~Tse and P.~Viswanath, \emph{Fundamentals of Wireless Communcation}.\hskip
  1em plus 0.5em minus 0.4em\relax Cambridge University Press, 2005.

\bibitem{Kushner2004Convergence}
H.~J. Kushner and P.~A. Whiting, ``Convergence of proportional-fair sharing
  algorithms under general conditions,'' \emph{IEEE Transactions on Wireless
  Communications}, vol.~3, no.~4, pp. 1250--1259, July 2004.

\bibitem{Stolyar2005on}
A.~L. Stolyar, ``On the asymptotic optimality of the gradient scheduling
  algorithm for multiuser throughput allocation,'' \emph{Operations Research},
  vol.~53, no.~1, pp. 12--25, January 2005.

\bibitem{Palomar2005Practical}
D.~P. Palomar and J.~R. Fonollosa, ``Practical algorithms for a family of
  waterfilling solutions,'' \emph{IEEE Transactions on Signal Processing},
  vol.~53, no.~2, pp. 686--695, February 2005.

\end{thebibliography}

\end{document}